\documentclass{elsarticle}
\usepackage{color,graphicx,url}
\usepackage{subfigure}

\usepackage{listings}
\journal{Parallel Computing}
\begin{document}
\begin{frontmatter}
  \title{Chunks and Tasks: a programming model for parallelization of dynamic algorithms}
  \author[label1]{Emanuel H. Rubensson}
  \ead{emanuel.rubensson@it.uu.se}
  \author[label1]{Elias Rudberg}
  \ead{elias.rudberg@it.uu.se}
  \address[label1]{Division of Scientific Computing, Department of Information Technology, Uppsala University, Box 337, SE-751 05 Uppsala, Sweden}


\begin{abstract}

We propose Chunks and Tasks, a parallel programming model built on
abstractions for both data and work. The application programmer
specifies how data and work can be split into smaller pieces, chunks
and tasks, respectively. The Chunks and Tasks library maps the
chunks and tasks to physical resources. In this way we seek to combine
user friendliness with high performance.
An application programmer can express a parallel algorithm using a few
simple building blocks, defining data and work objects and their
relationships. No explicit communication calls are needed; the
distribution of both work and data is handled by the Chunks and Tasks
library. 
This makes efficient implementation of complex applications that
require dynamic distribution of work and data easier.
At the same time, Chunks and Tasks imposes restrictions on data access
and task dependencies that facilitates the development of high
performance parallel back ends.
We discuss the fundamental abstractions underlying the programming
model, as well as performance and fault resilience considerations.  We
also present a pilot C++ library implementation for clusters of
multicore machines and demonstrate its performance for sparse blocked
matrix-matrix multiplication.

\end{abstract}

\end{frontmatter}

\section*{Submitted manuscript}

This manuscript was submitted to Parallel Computing (Elsevier) for the
special issue devoted to the PMAA 2012 conference.

\section{Introduction}
Parallel computing can be difficult.  In order to have a functioning
parallel program, one first needs to find algorithms that can be
executed in parallel. Then, the algorithms must be expressed in such a
way that the parallelism is exposed so that separate, independent
parts can be identified. Furthermore, the work must be distributed
among the available processors/nodes and data required by each
processor must be communicated as needed.  We present in this article
a parallel programming model intended to facilitate the above process
of developing parallel programs.

Our aim is to achieve scalability up to any number of processors, on
distributed memory machines and possibly on heterogeneous computers, to
handle not only algorithms with static distribution of work and data
but also hierarchic, recursive algorithms where communication patterns
are not known beforehand, and may change dynamically. Fault tolerance
is also important, especially considering the trends in modern
supercomputers, where hardware errors are unavoidable in any large
parallel calculation.

\subsection{Parallel programming models}
Much research has been devoted to parallel programming models. To give
some background to the design choices made while developing our new
interface, we below mention some of the more well-known previous
models. A programming model usually exists in the form of a tool,
e.g. a library or language, that supports program development within
the model. A crucial point here is the interface separating the
concerns of the application programmer from the inner workings of the
library or language.

In message passing programming models (e.g. MPI and PVM) the
application programmer has to decide how the data should be
distributed and use explicit communication to make sure that the
needed data is available whenever a task is to be performed.  This
gives the programmer control, but a drawback is that implementation of
complex algorithms, for example requiring dynamical load balancing, is
difficult.

An alternative that makes it easier to implement complex algorithms is
to use some programming model where data is shared via a distributed
shared memory.
The Linda programming model is built on an associative logically
shared memory called a tuple space. In implementations of Linda for
distributed memory each processor manages a portion of the tuple
space.  There is in principle no way for the Linda implementation to
know in advance which processes that will access a particular
tuple. However, efficient implementations attempt to optimize the
distribution of data based on observation of the tuple traffic between
processes \cite{Carriero1994633}. In case of static traffic patterns
the run-time system can then quickly set up efficient communication
channels. However, in cases of dynamic algorithms with a varying
communication pattern, it is not of much help.

Languages based on a Partitioned Global Address Space (PGAS) such as
Unified Parallel C (UPC) \cite{UPC-Consortium-2005} and CoArray Fortran give the programmer
more control of the distribution of data. Using PGAS approaches, a
skilled programmer can avoid extensive communication by distributing
and accessing data in a good manner.  On the downside, moving more
responsibilities to the application programmer reduces the
possibilities for the library/language to help.

In recent years there has been a growing interest in task-based
programming models.  A multitude of tools have been developed
including Cilk \cite{cilk}, SMPSs \cite{smpss} and OmpSs \cite{ompss},
StarPU \cite{starpu}, and SuperGlue \cite{lic-tillenius}.
Most of them have initially been developed for multicore
architectures, in some cases with support for accelerators. However,
there has also been some efforts to apply task-based approaches to
clusters of computers.  Cilk-NOW is a variant of Cilk for networks of
workstations \cite{cilk-now}, StarPU-MPI is an extension of StarPU for
clusters of accelerator-enhanced machines \cite{starpu-mpi}, and OmpSs
has been implemented for clusters of GPUs as well \cite{ompss}.  The
DAGuE framework~\cite{DAGuE-2012} is an example of a task-based
approach that achieves high performance in dense linear algebra
operations.

In task-based programming models, the programmer writes the program in
terms of tasks, usually specifying dependencies in one way or another,
and a runtime engine schedules the tasks on the available
resources. The programmer is responsible for exposing parallelism, and
the task scheduler is responsible for mapping the work to physical
resources. 
However, typically, either all data is managed by one ``master'' node or the
application programmer has to supply the distribution of data.

\subsection{The present work}
As stated above, a key issue is the division of labor between the
application programmer and the inner workings of the parallel
library/language.  In the present work we adopt a task-based approach
for the distribution of work.  However, we also provide abstractions
to handle data. The idea is that the application programmer should be
responsible for dividing data into smaller pieces to allow for data
distribution, but not for the mapping of this data to physical
resources.  
The application programmer (the user) defines task
types and registers tasks for execution. Similarly, the user defines
chunk types and registers chunks for storage.

Thus, a key feature of the present work is that the application
programmer is relieved from the burden of providing the data
distribution. This makes parallel programming a lot easier, especially
for applications where dynamic data structures play an important
role. Examples of such applications include sparse matrix operations
where the nonzero pattern is unknown beforehand and adaptive mesh
refinement.  For this type of applications, it is often beneficial to
use hierarchic data structures coupled to recursive algorithms.  One
notable example is large scale electronic structure calculations where
dynamic hierarchic data structures and algorithms are used with
success \cite{Bowler-review}.  Although our abstractions do not impose
a particular layout of work or data, we have had and have a view to
make them work well for dynamic hierarchic algorithms and data
structures.

We refer to our parallel programming model as the ``Chunks and Tasks''
programming model \cite{chunks-and-tasks} and it is presented here in
the form of a C++ application programming interface.  A Chunks and
Tasks library implementation is essentially composed of two parts, a chunk
management system and a task scheduler, that are responsible for
mapping data and work, respectively, to physical resources.

The main focus of the work presented here has been on establishing the
application programming interface. In our view, defining the interface
is most important since it in some sense is an answer to the question
of how to draw the line between the concerns of the application
programmer and the concerns of the parallel library.  Furthermore,
once the interface is fixed, application programmers can work on
implementing parallel algorithms independently of any improvements in
internal Chunks and Tasks library implementations. Also, there may be
several different Chunks and Tasks library implementations that an
application programmer can switch between without need for any changes
in the application program source code.  To further enforce a clean
interface with the possibility to switch between different
libraries we currently have two different library implementations of
Chunks and Tasks, a serial and a parallel implementation.
 
The Chunks and Tasks interface is described in Section~\ref{sec:cht}.
In Section~\ref{sec:pilot-impl}, we describe a pilot library
implementation based on MPI and POSIX threads and present a few
benchmark calculations.  Section~\ref{sec:discussion} contains some
discussion including performance and fault resilience considerations.
Finally, a brief summary and outlook is given in
Section~\ref{sec:conclusion}.

\section{The Chunks and Tasks interface} \label{sec:cht}
Chunks and Tasks is a parallel programming model, presented in this
article in the form of a C++ application programming interface. The
interface is intended to make it easier to parallelize any method that
can be formulated as a dynamic hierarchic algorithm.  Being evident
from the name, the interface is built around two central concepts:
chunks and tasks.

\begin{figure}
  \begin{center}
    \includegraphics[width=\textwidth]{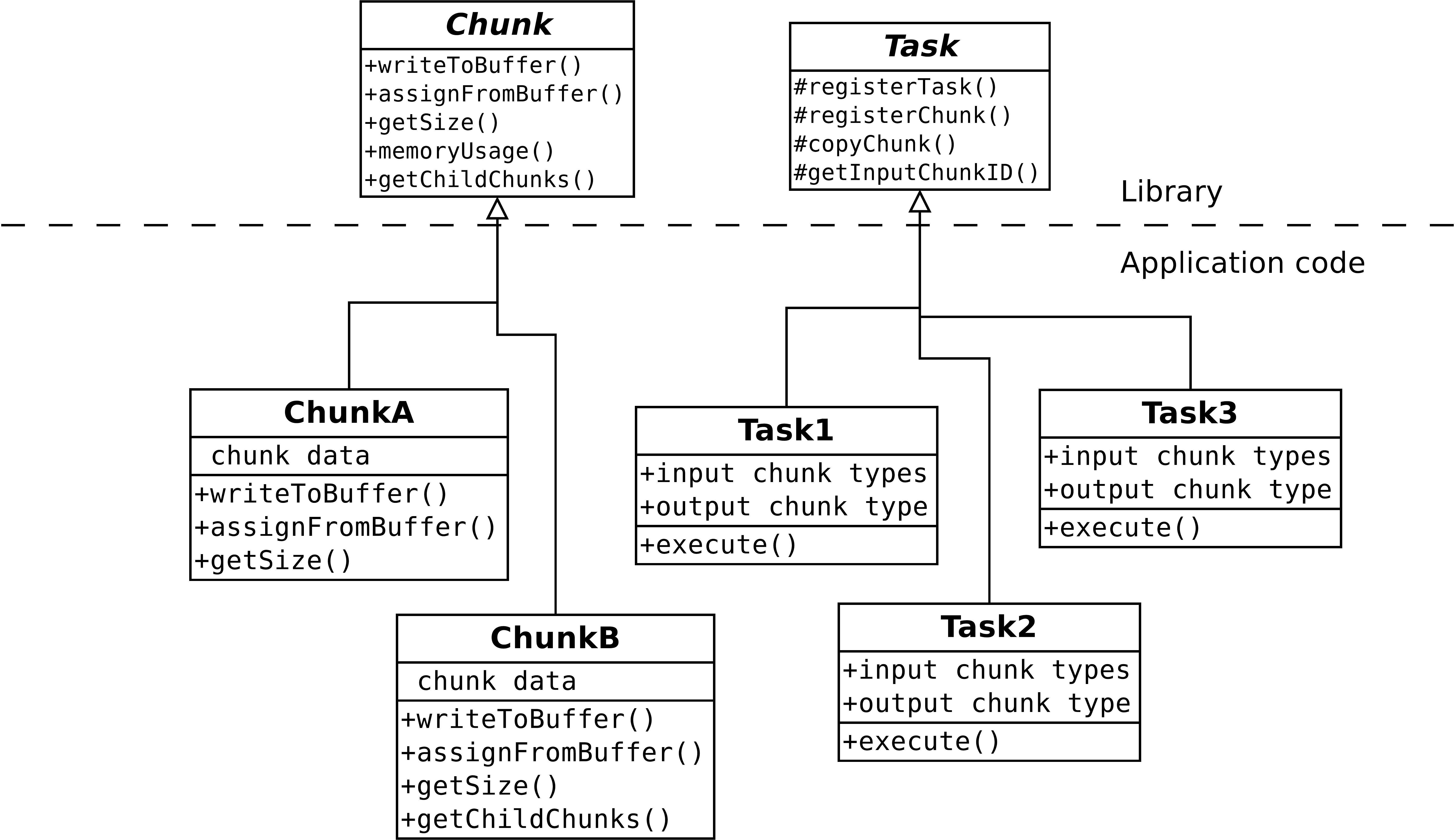}
  \end{center}
  \caption{The Chunks and Tasks interface consists of two base
    classes, Chunk and Task, from which the programmer derives data
    and task types for the application.  \label{fig-cht-classes}}
\end{figure}

The user defines chunk classes describing objects encapsulating pieces
of data. A chunk is registered by passing the control of the chunk
object to the Chunks and Tasks library. In return the user
receives a chunk identifier that later can be used to
specify data dependencies.  The user also defines task classes
describing work to be performed.  A task type is defined by a number of
input chunk types, the work to be performed and a single output chunk
type.
The relationships between the Chunk and Task base classes and user
defined classes are shown in Figure~\ref{fig-cht-classes}.

In this section, we describe the abstractions that are the basis of
the Chunks and Tasks interface.  In \ref{example-code} we provide a
concrete example program that uses Chunks and Tasks to compute a
Fibonacci number.

\subsection{The chunk abstraction} 

To define a chunk type you need to inherit from the Chunk base class
and define a few member functions. Functions to pack/unpack the
chunk to/from a given buffer are mandatory. Those functions may for
example be used by a Chunks and Tasks library when chunks need
to be sent to other processes over a network or to write chunks to
secondary storage.

After the point where the chunk has been registered and the user has
gotten hold of the chunk identifier (\verb|cht::ChunkID|) for that particular chunk,
modification of the chunk is not allowed.  Although any data or
information may be stored in a chunk object, one anticipated use is to
store identifiers to other chunks which in turn store chunk
identifiers and so on giving rise to a hierarchic data structure.

To make it possible for the library to take responsibility for
managing such chunk hierarchies, the user should in such cases
implement a function \verb|getChildChunks| that returns a list of all
chunk identifiers stored in that chunk. The list of child chunks can
be used by a library implementation in different ways, for example to
destruct chunk hierarchies, to prefetch child chunks, or to
send child chunks together with a chunk.

\subsection{The task abstraction} 
Task types are derived from the Task base class.  The programmer has
to specify the input chunk types and an output chunk type. The work to
be performed is specified by supplying an \verb|execute| function which takes
read-only  chunks as input and returns a chunk or task
identifier.  The work to be performed often includes registration of
new tasks.  In a task registration the task type and input identifiers
are specified and a task identifier (\verb|cht::TaskID|) is
returned. This task identifier may be used as input to subsequent task
registrations, thereby specifying dependencies between tasks. Since
each task produces a single output chunk, input for new tasks can be
specified using either chunk or task identifiers.

In our design, we have been careful to ensure that the task execution
can be performed without interruption, that is without waiting for
other tasks to finish or for communication. Access to chunks is only
allowed as input to tasks.  This means that it is not possible for the
Chunks and Tasks library to perform a task prior to having the
data needed for execution available. Besides the \verb|registerTask|
function, user defined task types inherit the \verb|registerChunk|,
\verb|copyChunk|, and \verb|getInputChunkID| functions from the Task
base class, see Figure~\ref{fig-cht-classes}.  All these functions
should be non-blocking.  Note that chunk registration and copy may
trigger communication with remote processes. However, the calling
thread does not need to wait for this communication, as will be
explained in Section~\ref{sec:task_scheduler_service}.

Task identifiers of any previously registered tasks may be specified as
input to a task.  Similarly to \cite{task-unified}, it is thus
possible to structure an algorithm both in a recursive divide and
conquer style and with dependencies on any ancestor. However, contrary
to \cite{task-unified}, since chunks are read-only, neither the
library developer nor the application programmer need to worry about
races for shared data access. It is also impossible to end up in a
dead-lock due to mistakes in application code since all tasks directly or
indirectly only depend on read-only data.

\subsection{The main program}
A calculation performed using Chunks and Tasks is
started from a standard serial C++ main program written by the application programmer. In
a similar way as in a task execution, the user constructs chunks to be
used as input to tasks and registers tasks to be executed by the
Chunks and Tasks library.

Often, there is a working serial implementation for a particular
application. It is then desirable to be able to parallelize the most
time consuming part(s) without rewriting the whole code. We allow the
main program to remain a serial code with the Chunks and Tasks
parallelization employed only for selected parts.

\section{A pilot library implementation} \label{sec:pilot-impl}
Unlike most MPI and PGAS programs our pilot Chunks and Tasks library implementation does not
use a Single Program Multiple Data (SPMD) style. Instead, the program starts as a serial program and at the Chunks and Tasks
start-up the worker processes are spawned. These processes execute a
worker program provided by the Chunks and Tasks library.
Each worker starts a set of services, including a chunk service and a
task scheduler service. Each service uses its own MPI communicators.

Our present library implementation uses the \verb|MPI_Comm_spawn|
function in the MPI 2 standard.  If a fully compliant MPI 2 library is
not available, all processes can be started simultaneously and then
assume the roles of parent and workers. However, we believe that the
spawn functionality is valuable since it provides the possibility of
dynamically adapting the amount of resources used by a calculation.

\subsection{Chunk service}
When the chunk service starts, on each worker a thread is created that
is responsible for listening for MPI messages from other workers
issued by the chunk service on those workers. Other
workers may request to get, delete, or copy a chunk.

When a chunk is registered, the size of the chunk is known, and it is
therefore possible to store the size of the chunk in the chunk
identifier. Besides obvious practical benefits for chunk
communication, this information may for example be used in parametric
models to estimate task execution times.
The chunk identifier also contains the MPI rank of the worker where
the chunk is stored, so when a chunk needs to be fetched we directly
know to which worker the request should be sent.
New chunks are by default assigned to the local worker, so
that no communication is needed to register new chunks.

In the chunk identifier is also stored a chunk \emph{type} identifier.  Given a chunk
type identifier and the serialized chunk data, a chunk object can be
reconstructed on another worker. A chunk factory constructs a chunk
object of correct type given the chunk type identifier, and a call to
the \verb|assignFromBuffer| function completes the reconstruction.

The chunk service also implements chunk cache functionality: when a
remote chunk has been fetched over the network, it is kept in memory
so that we do not need to fetch it again if the chunk is requested
multiple times. Cached chunks are purged from memory in a least
recently used fashion.

\subsection{Task scheduler service} \label{sec:task_scheduler_service}
Our task scheduler is based on work stealing. The calculation is
initiated by the parent process sending the mother task to one of the
workers. This worker begins to execute tasks in a depth first fashion,
working its way down in the task hierarchy. Whenever a worker is out
of work, it attempts to steal work from another worker chosen at
random. In order to achieve as much parallelism as possible, tasks are
always stolen as high up in the task hierarchy as possible.

When a task is stolen, the stealing worker has to reconstruct the
task. As in the chunk reconstruction, the task identifier includes a
task type identifier that is used in a call to a task factory which
constructs a task object of the correct type.

When the task scheduler service starts, several threads are created on
each worker process. As for the chunk service, one thread is responsible for
listening for MPI messages from the task scheduler service running on
other processes.  These messages can include task steal attempts or
task information.  Another thread is fetching data for tasks pending
for execution, to achieve overlap of communication and computation. 
A number of threads execute tasks.

\subsubsection{The task transaction} \label{sec:transaction}
The outcome of a task execution consists of the output chunk or task
identifier and the effect of calls to \verb|registerChunk|,
\verb|copyChunk|, and \verb|registerTask|. It is possible to implement
the effect of these operations immediately. However, in our pilot
implementation, the aggregate effect of a task execution is performed
in a single transaction at some point after the execution of the task
has finished, in a way similar to the return transaction technique
proposed by Blumofe and Lisiecki~\cite{cilk-now}.  This is
accomplished by accumulating all the information needed for the
transaction during the task execution.

In this way, the task execution can be performed without interruption,
even if a call to for example \verb|copyChunk| would result in
communication with a remote process.  Other benefits include the
possibility to do speculative execution of tasks.

\subsubsection{Speculative task execution}
On each worker, several threads execute tasks.  In order to have
efficient task stealing between workers, we need to avoid unrolling
additional branches of a task hierarchy when there is a lot of work in
a branch that is already about to be unrolled.  This could be achieved
by only allowing one thread at a time to execute non-leaf tasks,
i.e.~tasks that register child tasks. The problem is that, in
principle, the only way to know if a task is a leaf task or not is to
run \verb|execute| for that task. To circumvent this problem we use what we
refer to as \emph{speculative task execution}.  Tasks are executed
speculatively in the sense that it is not known, at the time of
execution, if the task transaction will be performed.  For tasks that
turn out to be leaf tasks, the task transaction is performed
immediately. For non-leaf tasks, task transactions are only allowed
one at a time. In this way we avoid undesired unrolling of several
branches of a task hierarchy at the same time.  An executed task for which the
transaction has not been performed, can still be stolen, 
ensuring that stealing still occurs as high as possible in the task hierarchy.

\subsubsection{Side effects of task execution}
All effects of a task are collected in the transaction for that
task. If the transaction is considered as the outcome of a task, then
the task is completely free from side effects. However, if the output
chunk or task identifier is considered as the outcome of the task,
then the effect of calls to \verb|registerChunk|, \verb|copyChunk|,
and \verb|registerTask| during task execution should strictly speaking
be considered as side effects. However, the ultimate effect of such
calls is the creation of chunks whose identifiers from user code can
only be reached through the output identifier.  This means that
dropping the output identifier without further notice would only
result in a chunk leak.  This has important implications for fault
resilience as will be discussed in Section~\ref{sec:Fault-resilience}.

\subsection{Test calculations}
To test our pilot implementation we have written a test program that
implements matrix-matrix multiplication for hierarchic block-sparse
matrices using the Chunks and Tasks programming model. 
The matrices are represented by quad-trees of chunk
identifiers. At the lowest level, each nonzero submatrix is
represented by a regular full matrix. At higher levels, four chunk
identifiers are stored referring to submatrices at the next lower
level. If a submatrix is zero it is represented by the special chunk
identifier \verb|cht::CHUNK_ID_NULL|. This implementation corresponds
to the hierarchic block-sparse matrix data structure in~\cite{m-rrs}
now expressed using the Chunks and Tasks interface.

The matrix-matrix multiplication is implemented using three task
types; one for matrix-matrix multiplication, one for matrix-matrix
addition, and one to construct a matrix from the chunk identifiers of
the four submatrices. Sparsity is handled by checking for
\verb|cht::CHUNK_ID_NULL|.  The same matrix-matrix multiplication
implementation was used for all test calculations discussed below, for
both dense and block-sparse cases.

We performed test calculations on the Tintin cluster at the UPPMAX
computer center. Each compute node is a dual AMD Bulldozer compute server
with two 8-core Opteron 6220 processors running at 3.0 GHz, with 64 GB
of memory per node. The nodes are interconnected with a 2:1
oversubscribed QDR InfiniBand fabric.  Since the 16 cores on each node
share 8 256-bit fused multiply-add (FMA) units, each node can at best
perform 8*4 double precision multiply-add operations per cycle. Given
the clock speed of 3.0 GHz, this gives a theoretical peak performance
of 96 GFlop/s per node.
However, one cannot expect to reach the theoretical peak performance
in practice. Since we used the AMD Core Math Library (ACML) for
matrix-matrix products at the lowest level, the ACML performance for
large dense matrices can be seen as a practical peak performance
limit. We refer to this as ``ACML peak'', which we computed from dense
matrix-matrix multiplication with ACML for matrix size 12000 $\times$
12000 using all cores on the node. This gave an ``ACML peak'' value of
80.171 GFlop/s per node.

\begin{figure}
  \begin{center}
    \includegraphics[width=0.5\textwidth]{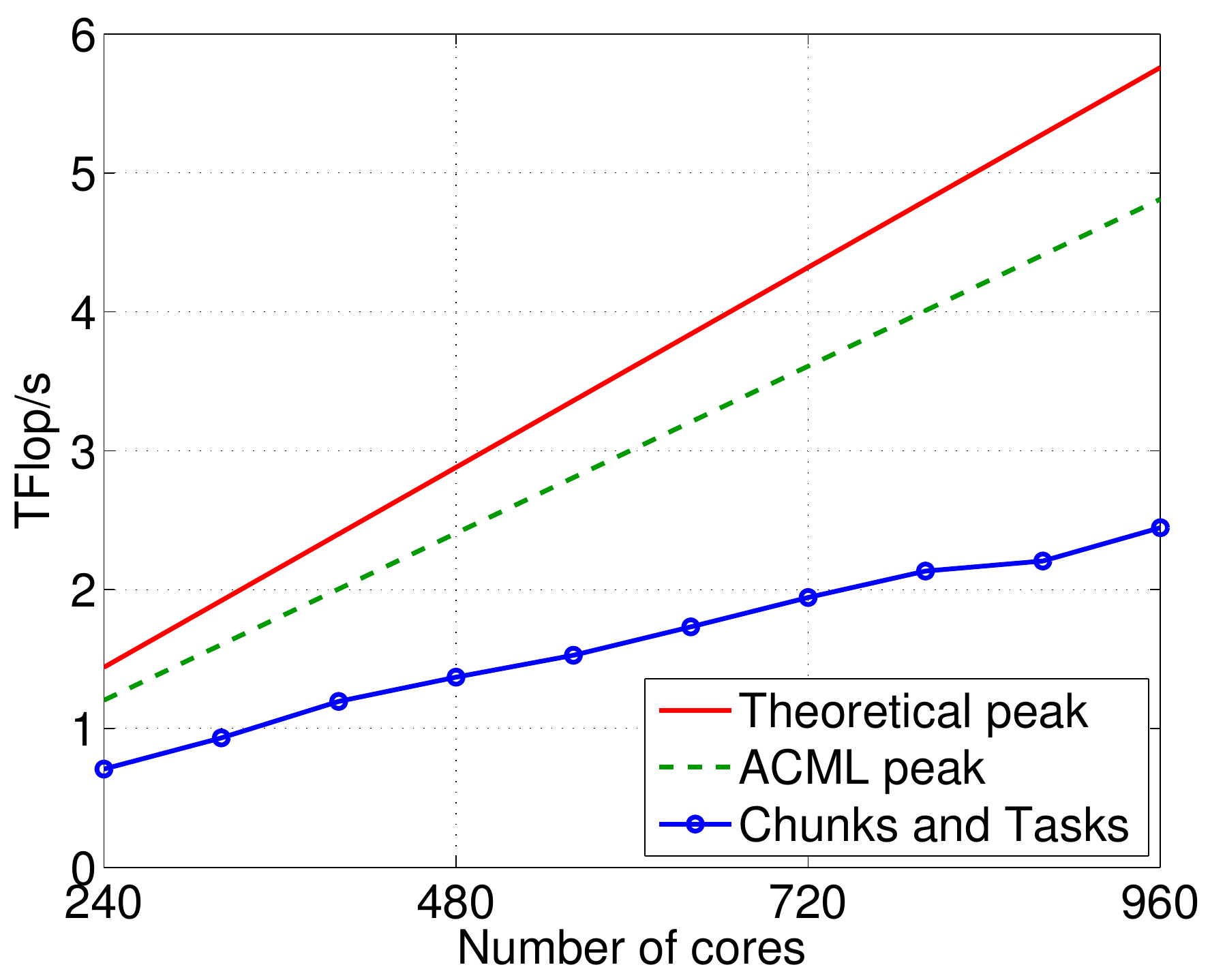}
  \end{center}
  \caption{Strong scaling for dense matrix-matrix multiplication for
    matrix size 60000 $\times$ 60000
\label{fig-strong-scaling}}
\end{figure}

\begin{figure}
  \begin{center}
    \includegraphics[width=0.5\textwidth]{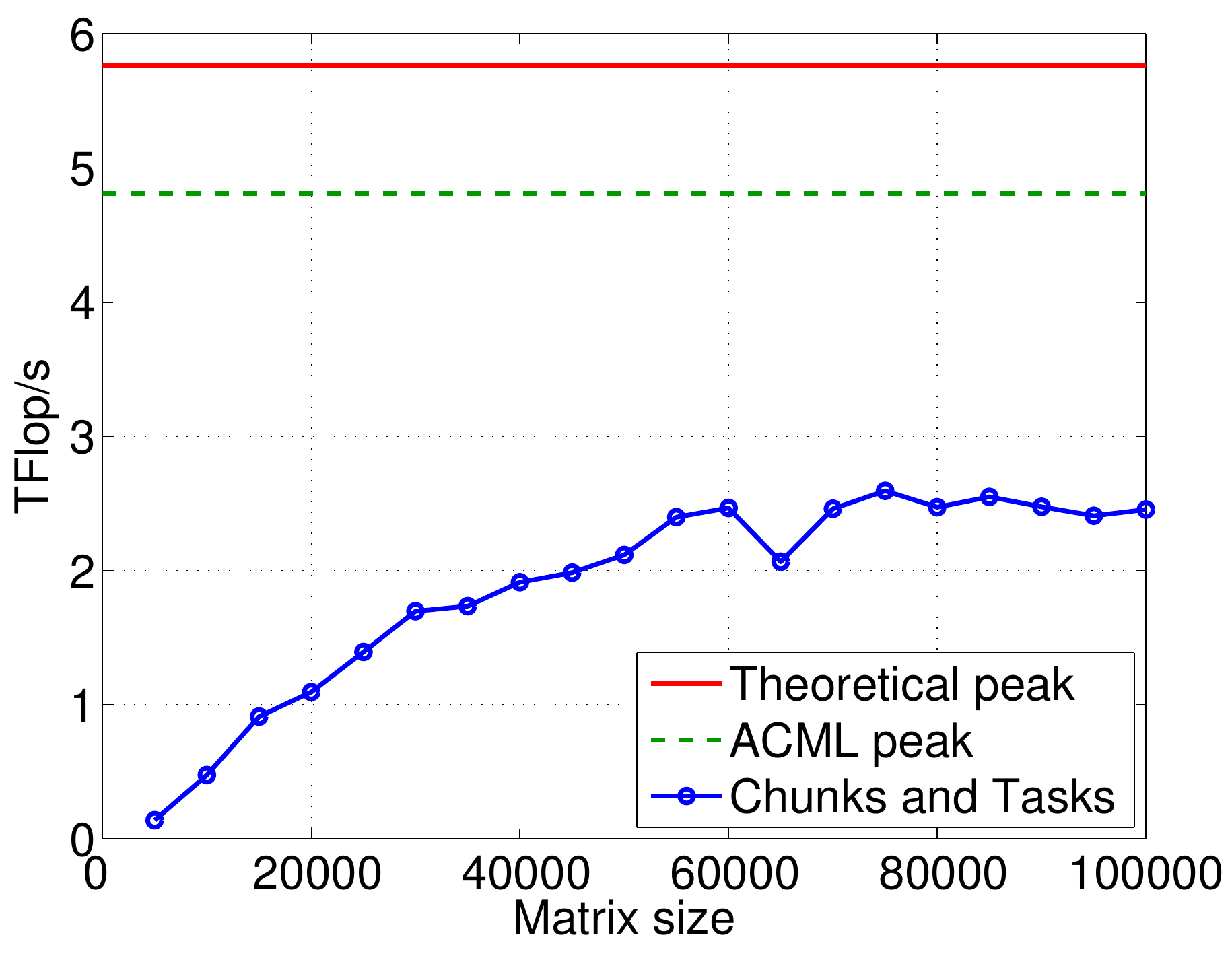}
  \end{center}
  \caption{Dense matrix-matrix multiplication performance when running
    on 60 nodes (960 cores), for varying matrix size.
\label{fig-problem-size-scaling}}
\end{figure}

In our test calculations the Chunks and Tasks library used one MPI
process per node, with 15 worker threads per process.
Figure \ref{fig-strong-scaling} shows strong scaling behavior for
dense matrix-matrix multiplication for matrix size 60000 $\times$ 60000 when
increasing the number of nodes from 15 to 60.
Figure \ref{fig-problem-size-scaling} shows the performance when
increasing the problem size for dense matrix-matrix multiplication
when running on 60 nodes.
The submatrix size at the lowest
level was 1000 $\times$ 1000.

\begin{figure}
  \center
  \subfigure[Timings]{
    \includegraphics[width=0.49\textwidth]{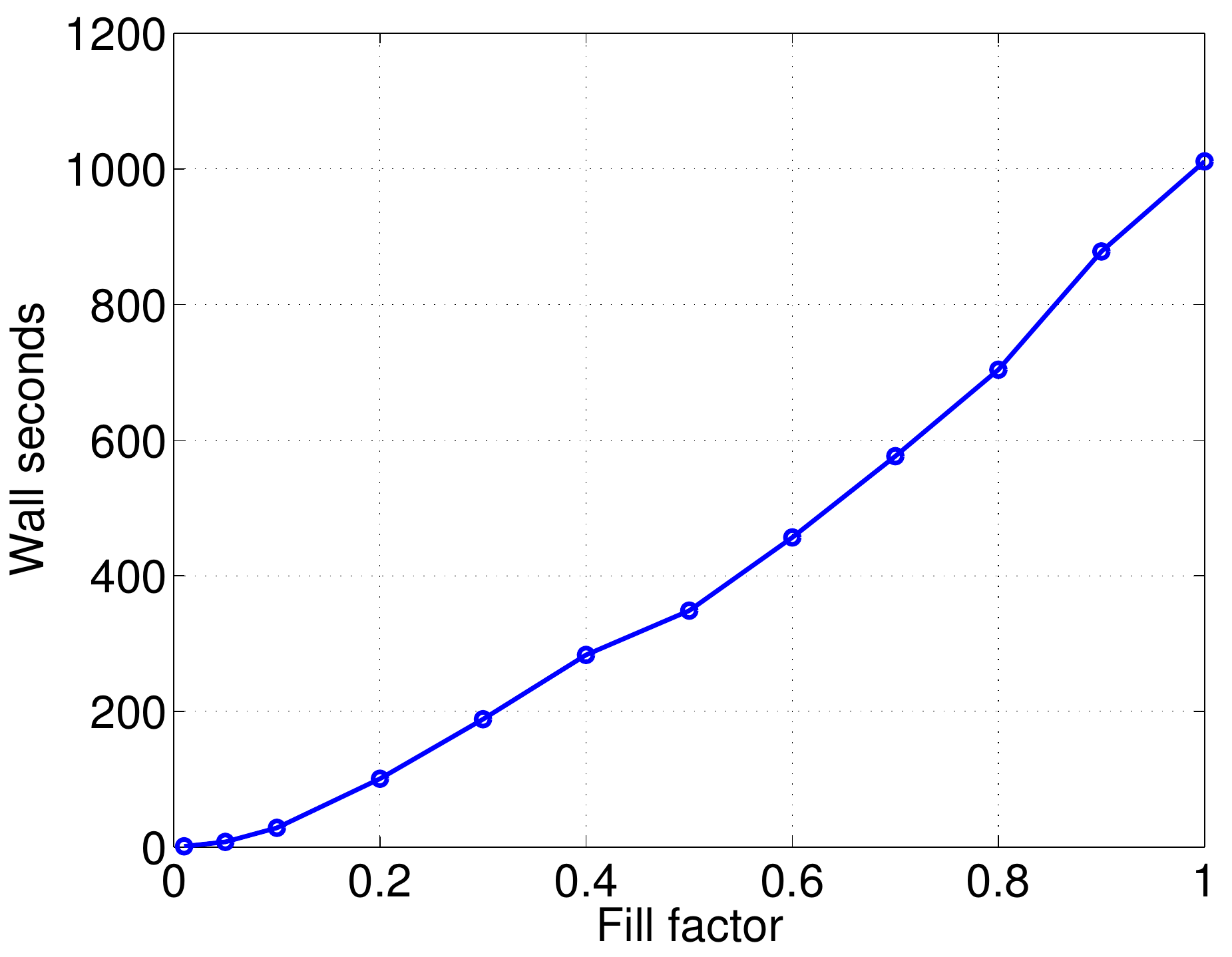}
  }
  \subfigure[Nonzero pattern]{
    \includegraphics[width=0.41\textwidth]{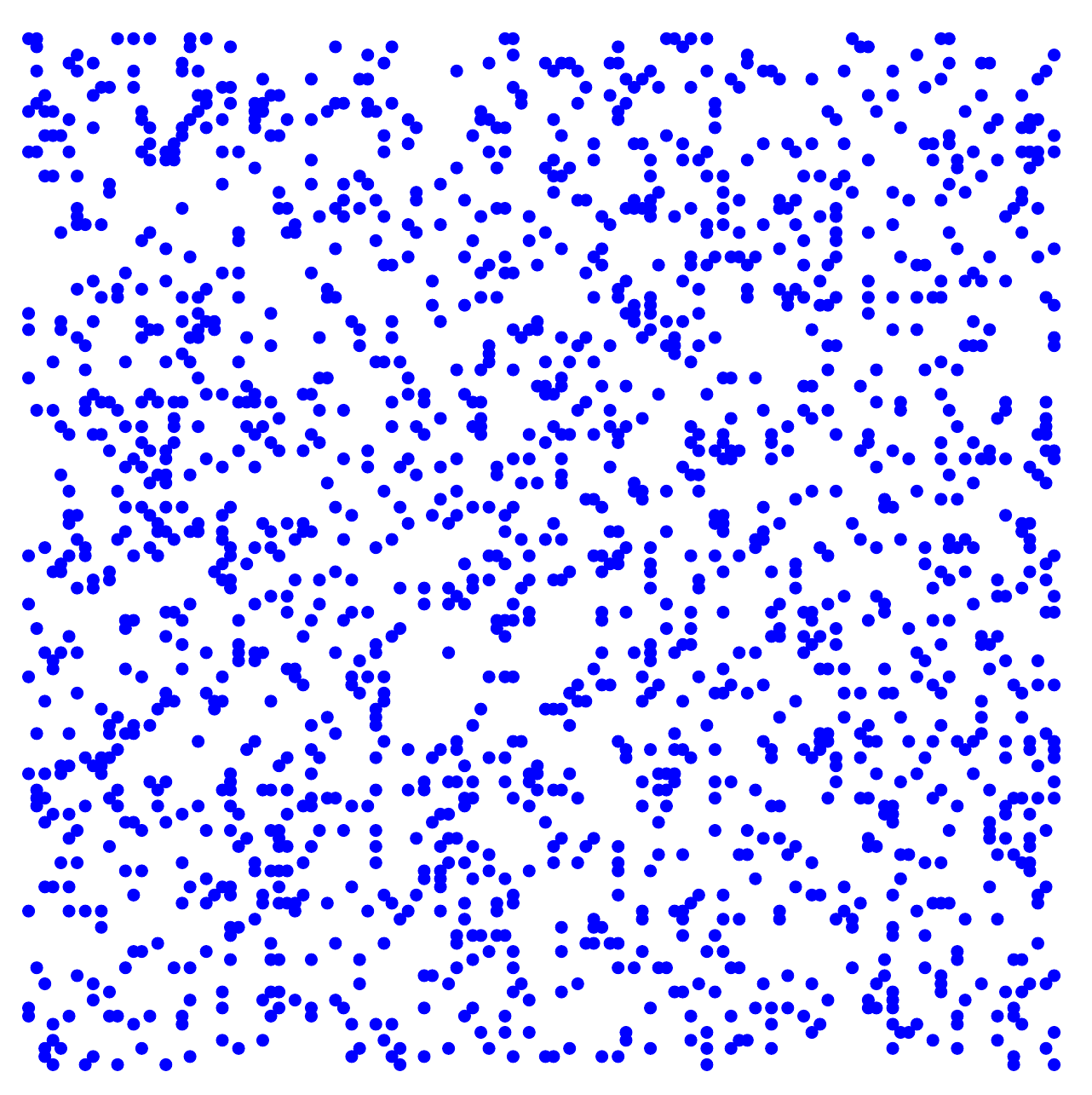}
  }
  \caption{Results from block-sparse matrix-matrix multiplication test
    runs with matrix size $128000 \times 128000$ and varying sparsity,
    running on 60 nodes (960 cores) on the Tintin cluster.  Panel (a):
    Wall time as a function of fill factor.  The fill factor is the
    fraction of elements that are nonzero, so the rightmost point in
    the left panel corresponds to the dense matrix case.  Panel (b):
    Nonzero pattern for the case with fill factor 0.1, i.e. 10\%
    nonzero elements. Each dot represents 1 million nonzero elements.
    \label{fig-varying-fill-factor}}
\end{figure}

\begin{figure}
  \begin{center}
    \includegraphics[width=0.5\textwidth]{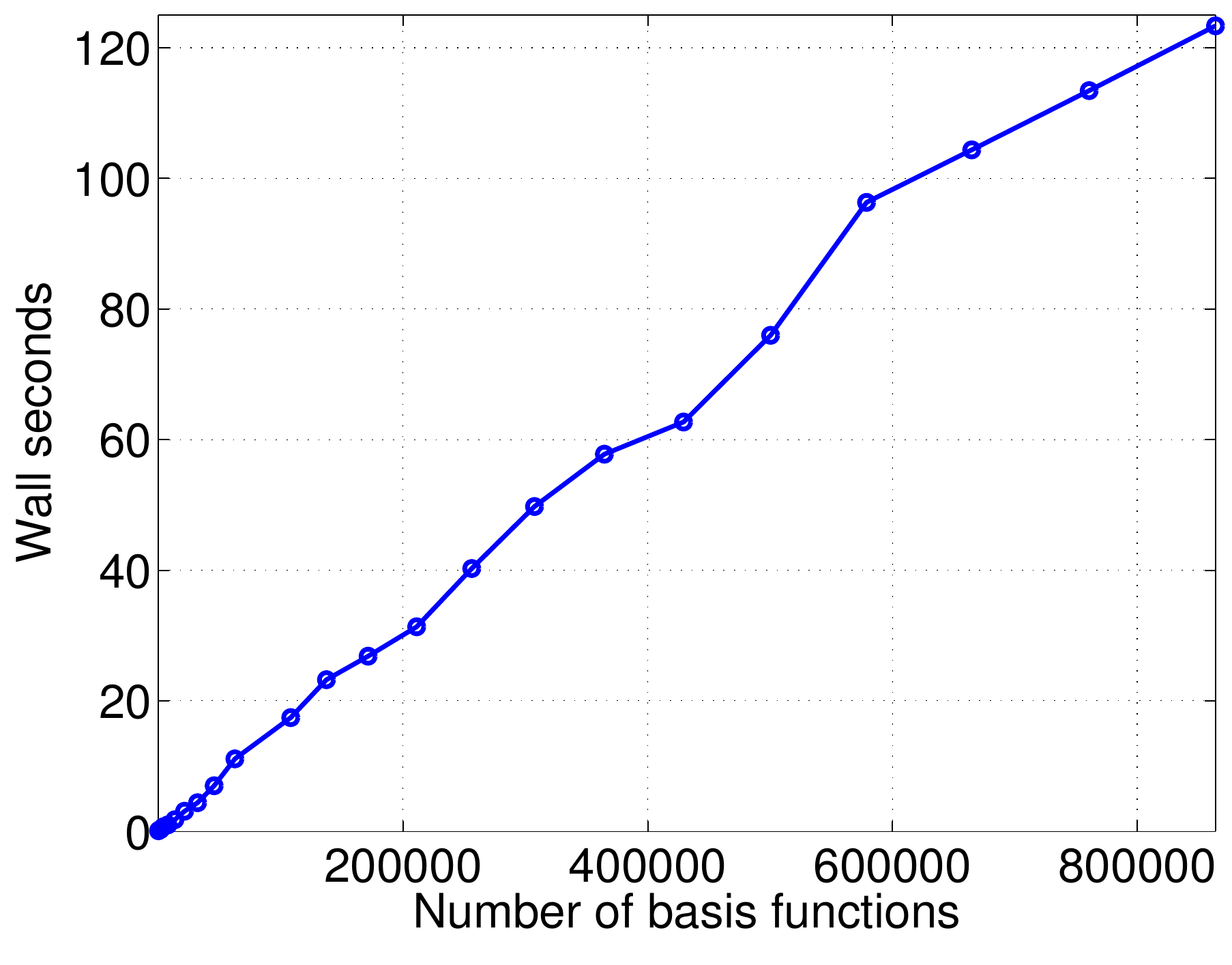}
  \end{center}
  \caption{Timings for computation of the square of the overlap matrix
    for water clusters of varying size, running on 60 nodes (960
    cores) on the Tintin cluster. The Gaussian basis set STO-3G was
    used. The largest water cluster consisted of 123457 water
    molecules, corresponding to 864199 basis functions.
    \label{fig-water-clusters-scaling}}
\end{figure}

Figure~\ref{fig-varying-fill-factor} shows results of matrix-matrix
multiplication for block-sparse matrices with random sparsity patterns
of varying fill factor.  Given the specified fill factor the nonzero
submatrices were uniformly randomly distributed over the matrix. The
figure shows how the wall time for the matrix square operation
decreases with increasing sparsity.

As an initial test of the usefulness of the Chunks and Tasks
programming model for a specific application, linear scaling
electronic structure calculations, we have parallelized the overlap
matrix computation in the Ergo quantum chemistry program\cite{m-ergo}
using Chunks and Tasks.

The Ergo program performs electronic structure calculations using
Gaussian basis sets, where typically a number of basis functions are
centered on each atom. A necessary first step in such a computation is
to compute the overlap matrix $S$. To parallelize this procedure we
have used a hierarchic representation of the basis set, where each
part of the hierarchy contains basis functions located in a particular
part of space. In the same way as for the sparse matrix
representation, higher levels in the hierarchy contain chunk
identifiers referring to basis set descriptions at lower levels. Using
such a hierarchic basis set description, it is straightforward to
implement tasks to compute the overlap matrix.

One of the most performance-critical parts in linear scaling
electronic structure calculations is matrix-matrix multiplication; in
particular, methods for computation of the density matrix usually rely
on repeated sparse matrix-matrix multiplication. We have tested the
performance of our matrix-matrix multiplication implementation based
on Chunks and Tasks by computing the square of the overlap matrix for
water clusters of varying
size. Figure~\ref{fig-water-clusters-scaling} shows timings for the
computation of $S^2$ when running on 60 nodes.
As expected, the computational time scales roughly linearly with the
water cluster size.

The calculations shown in Figure~\ref{fig-water-clusters-scaling} were
done for water clusters of up to 123457 water molecules, corresponding
to 864199 Gaussian basis functions using the STO-3G basis set. A
submatrix size of 500 $\times$ 500 was used at the lowest level. The overlap matrix
was truncated so that the Frobenius norm of the error matrix was
smaller than $10^{-5}$. For the largest water cluster, the overlap
matrix then contained 1.20 \% nonzero elements, and the product $S^2$
contained 5.34 \% nonzero elements. 
This corresponds to a total memory requirement of about 391 GB to store $S$ and $S^2$.

\section{Discussion}\label{sec:discussion}

\subsection{Responsibility for data distribution}

The Chunks and Tasks interface allows the application programmer to
define distributed data objects by creating chunks that refer to other
chunks through their chunk identifiers. The programmer does not need to
specify where the chunks should be stored. This gives great
flexibility, especially when implementing dynamic hierarchic
algorithms where the structure of data to be created is not known
beforehand. 

Thus, compared to other parallelization approaches where the data
distribution is explicitly specified by the programmer, the Chunks and
Tasks approach moves more responsibility to the library, letting the
programmer focus on algorithm development and to exposing parallelism
by making sure distributed objects are represented by trees of chunks.

Relieving the programmer of the responsibility for data distribution
means additional challenges when implementing a Chunks and Tasks
library. However, as demonstrated in Section~\ref{sec:pilot-impl},
high performance can still be achieved.  Chunks and Tasks library
implementations have great freedom in how data is managed and where
tasks are run depending on the required input data.  Our pilot
implementation demonstrates one possible way to manage data, but many
others are certainly possible. Importantly, application programs
implemented using the Chunks and Tasks interface can remain unchanged
and still benefit from any future improved library implementations.

\subsection{Restrictions and performance}

The Chunks and Tasks programming model implicitly imposes certain
restrictions on how work and data is handled in an application
program. Chunks are read-only. Chunks and tasks are identified by
their identifiers provided by the Chunks and Tasks library upon
registration. It is therefore not possible to register tasks with
dependencies on tasks that have not yet been registered. It is also
not possible to create task hierarchies with dependencies across
branches.

This stands in contrast to the Linda programming
model~\cite{Carriero1994633} where tuples are accessed associatively,
with none of the above restrictions.  However, the restrictions
imposed by Chunks and Tasks make implementation of efficient parallel
Chunks and Tasks libraries feasible.  Chunk cache coherence is not an issue.  There is
no need to resolve dependencies on remote tasks. Furthermore, an
efficient library can make use of the chunk identifiers to make
data available efficiently, or choose to run tasks where the data is
located.

Another feature made possible due to the restrictions imposed by the 
interface is efficient implementation of the \verb|copyChunk| function.
From a user perspective, \verb|copyChunk| takes a chunk identifier,
copies the chunk, and returns an identifier for the new chunk (the
copy).

However, a library implementation of \verb|copyChunk| can take
advantage of that chunks are read-only.  A copy is then performed by
creating a new chunk identifier that refers to the same chunk.  The
library counts all chunk identifiers that refer to the same chunk and
delete the whole chunk hierarchy represented by that chunk only when
the last copy of the chunk is being deleted.  Thus, although the copy
is actually a \emph{shallow copy}, from a user perspective it should
be considered as a \emph{deep copy}.

So what are the practical implications of this?  Consider, for
example, an adaptive mesh refinement algorithm where the mesh is
represented by a chunk hierarchy and say that we want to refine a
local portion of the mesh.  Since chunks are read-only one might think
that such a local refinement would incur a replication of the entire
mesh. However, thanks to the copy chunk functionality, only the
refined region needs to be reconstructed.

\subsection{Fault-resilience}\label{sec:Fault-resilience}
A fault-resilient program is able to continue executing in case of a
failure, e.g.~a worker crash.
Communication systems such as MPI are not able to single-handedly,
without intervention by the application program, handle faults
\cite{fault-tolerant-GroppFall}. The reason is that, in this
programming model, the distribution of data and the program to be
executed by each process is the concern of the application programmer
and not the MPI library.  The best thing the communication system can
do is to gracefully report failures to the application program which
then may take appropriate measures \cite{fault-tolerant-dongarra}.

One approach to achieve fault-tolerance at the application level is to
maintain a shadow copy of all data that is needed to recover in case
of a process failure \cite{fault-tolerant-Vishnu}.  Global Arrays
includes functionality to control the mapping of a global array onto
the processes. This makes it it possible for the user to make sure
that primary and shadow copies do not overlap
\cite{fault-tolerant-Krishnamoorthy}.

Similarly to \cite{cilk-now}, it should with Chunks and Tasks be
possible to achieve resilience to faults at the Chunks and Tasks
library level.  Provided that a fault-resilient Chunks and Tasks
library is used, a conforming Chunks and Tasks application will
automatically be fault-resilient. This can be seen as a consequence of
the Chunks and Tasks not being a SPMD programming model; worker
processes are run by the library rather than by the application.

In order for the Chunks and Tasks library to be resilient, it
has to include mechanisms for chunk backup and re-execution of failed
tasks. Since tasks in Chunks and Tasks do not have critical side
effects, a fail-safe library does not need to deal with failed
tasks that have produced partial output, as in for
example~\cite{fault-tolerant-Dinan}. If a task fails, one can simply
re-execute it. Recovery of chunks lost due to a process failure can be
achieved by storing a shadow copy on another process as described
above.

\section{Concluding remarks}\label{sec:conclusion}
We have presented the Chunks and Tasks programming model, our answer
to the question of how to draw the line between the concerns of an
application programmer and the concerns of a parallel library or
language.  Our philosophy is that application programmers should focus
on parallel algorithm development and on \emph{exposing} parallelism
in both data and work.

This makes the development of parallel programs easier, in particular
for applications that require dynamic distribution of both work and
data.  At the same time, Chunks and Tasks imposes restrictions on data
access and task dependencies that make it possible to implement Chunks
and Tasks libraries with high performance. Furthermore, it is possible
to achieve fault resilience at the Chunks and Tasks library level,
which derives from the fact that the library is responsible for
mapping of both work and data to physical resources.

We see before us that parallelization tools building on the Chunks and
Tasks programming model will expand the applicability of high
performance parallel computing to an important class of applications
that require dynamic handling of work and data.

\section*{Acknowledgment}
Stimulating discussions at the PSCP colloquium at Uppsala University as well as support from
the Swedish Research Council under Grant No.~623-2009-803 and from the
Swedish national strategic e-science research program eSSENCE are
gratefully acknowledged.  The calculations were performed on resources
provided by the Swedish National Infrastructure for Computing (SNIC)
at Uppsala Multidisciplinary Center for Advanced Computational Science
(UPPMAX).

\appendix
\section{Example usage}\label{example-code}
\lstset{basicstyle=\ttfamily\small, 
  language=[ISO]C++, morekeywords={iostream}, 
  keywordstyle=\color[rgb]{0.2,0.5,0.2},
  showspaces=false, showstringspaces=false,
  frame=lines, 
  numbers=left, firstnumber=1, numberstyle=\tiny }

\begin{lstlisting}[caption=Main program]
#include <iostream>
#include "chunks_and_tasks.h"
#include "Fibonacci.h"
int main() {
  int n = 13;
  cht::start();
  cht::ChunkID cid_n = cht::registerChunk(new CInt(n));
  cht::ChunkID cid_result = 
    cht::executeMotherTask<Fibonacci>(cid_n);
  // Get result.
  cht::shared_ptr<CInt const> result;
  cht::getChunk(cid_result, result);
  // Delete chunks.
  cht::deleteChunk(cid_n);
  cht::deleteChunk(cid_result);
  // Stop cht services
  cht::stop();
  std::cout << "The thirteenth Fibonacci number is " 
            << *result << std::endl;
  return 0;
}
\end{lstlisting}

\begin{lstlisting}[caption=CInt.h]
#include "chunks_and_tasks.h"
struct CInt: public cht::Chunk {
  // Functions required for a Chunk
  void writeToBuffer(char * dataBuffer, 
		     size_t const bufferSize) const;
  size_t getSize() const;
  void assignFromBuffer(char const * dataBuffer, 
			size_t const bufferSize);
  size_t memoryUsage() const;
  // CInt specific functionality
  CInt(int x_) : x(x_) { }
  CInt() { }
  operator int() const { return x; }
 private:
  int x; // The number itself
  CHT_CHUNK_TYPE_DECLARATION;
};
\end{lstlisting}

\begin{lstlisting}[caption=CInt.cc]
#include <cstring>
#include "CInt.h"

CHT_CHUNK_TYPE_IMPLEMENTATION((CInt));
void CInt::writeToBuffer(char * dataBuffer, 
			 size_t const bufferSize) const {
  if (bufferSize != getSize())
    throw std::runtime_error("Wrong buffer size to "
			     "CInt::writeToBuffer.");
  memcpy(dataBuffer, &x, sizeof(int));
}
size_t CInt::getSize() const {
  return sizeof(int);
}
void CInt::assignFromBuffer(char const * dataBuffer, 
			    size_t const bufferSize) {
  if (bufferSize != getSize())
    throw std::runtime_error("Wrong buffer size to "
			     "CInt::assign_from_buffer.");
  memcpy(&x, dataBuffer, sizeof(int));
}
size_t CInt::memoryUsage() const {
  return getSize();
}
\end{lstlisting}

\begin{lstlisting}[caption=Fibonacci.h]
#include "chunks_and_tasks.h"
#include "CInt.h"

struct Fibonacci: public cht::Task {
  cht::ID execute(CInt const &);
  CHT_TASK_INPUT((CInt));
  CHT_TASK_OUTPUT((CInt));
  CHT_TASK_TYPE_DECLARATION;
};
\end{lstlisting}

\begin{lstlisting}[caption=Fibonacci.cc]
#include "Fibonacci.h"

struct Add: public cht::Task {
  cht::ID execute(CInt const & n1, CInt const & n2);
  CHT_TASK_INPUT((CInt, CInt));
  CHT_TASK_OUTPUT((CInt));
  CHT_TASK_TYPE_DECLARATION;
};

CHT_TASK_TYPE_IMPLEMENTATION((Add));
cht::ID Add::execute(CInt const & n1, CInt const & n2) {
  CInt result_chunk = n1+n2;
  cht::ChunkID cid_result = 
    registerChunk(new CInt(result_chunk), cht::persistent);
  return cid_result;
}

CHT_TASK_TYPE_IMPLEMENTATION((Fibonacci));
cht::ID Fibonacci::execute(CInt const & n) {
  if(n < 2) 
    return copyChunk( getInputChunkID(n) );
  cht::ChunkID c1 = registerChunk( new CInt(n-1) );
  cht::ID t1 = registerTask<Fibonacci>( c1 );
  cht::ChunkID c2 = registerChunk( new CInt(n-2) );
  cht::ID t2 = registerTask<Fibonacci>( c2 );
  return registerTask<Add>(t1, t2, cht::persistent);
} // end execute
\end{lstlisting}

\end{document}